\documentclass[11pt]{article}
\usepackage[T1]{fontenc}
\usepackage[utf8]{inputenc}
\usepackage{amsmath,amssymb,amsthm,mathrsfs}
\usepackage[margin=1in]{geometry}
\usepackage{hyperref}

\newtheorem*{remark*}{Remark}

\title{An exact unstable mode of self-similar\\ Yang--Mills solutions
in $5+1$ dimensions}
\author{%
Piotr Bizo\'n\thanks{Email: \texttt{piotr.bizon@uj.edu.pl}}\\[1ex]
 Institute of Theoretical Physics,
Jagiellonian University, Krak\'ow
}

\date{}

\begin{document}
\maketitle

\begin{abstract}
Self-similar solutions play an important role in the formation
of singularities in the Yang--Mills equations. In five space
dimensions, these equations admit infinitely many self-similar
solutions with an increasing number of unstable modes. It was observed numerically in~\cite{B} that all excited profiles
share the least unstable eigenvalue $\lambda=5$.
We explain this remarkable spectral property by an elementary
construction of the corresponding eigenfunction.
\end{abstract}

\subsection*{Self-similar solutions}

The spherically symmetric Yang--Mills equation in five
space dimensions is
\begin{equation}\label{YM}
w_{tt}=w_{rr}+\frac2r w_r+\frac3{r^2}w(1-w^2).
\end{equation}
In terms of the similarity coordinates
\[
y=\frac r{T-t},
\qquad
\tau=-\log(T-t),
\]
where $T$ is the blowup time, equation \eqref{YM} becomes
\begin{equation}\label{similarity}
w_{\tau\tau}+w_\tau+2y w_{\tau y}
=
(1-y^2)w_{yy}
+\left(\frac2y-2y\right)w_y
+\frac3{y^2}w(1-w^2).
\end{equation}

By definition, self-similar solutions are independent of $\tau$:
$w(\tau,y)=u(y)$. The  profile $u$ therefore satisfies
\begin{equation}\label{profile-y}
(1-y^2)u''+\left(\frac2y-2y\right)u'
+\frac3{y^2}u(1-u^2)=0,\qquad 0<y<1.
\end{equation}
It was proved in \cite{B} that there exist infinitely many
 smooth profiles $u_n$, $n=0,1,2,\ldots$,
where $u_n$ has exactly $n$ zeros on $(0,1)$.
Their endpoint expansions are
\begin{align}
u_n(y)&=1-a_ny^2+O(y^4),&&y\to0,\label{center-y}\\
u_n(y)&=b_n(1-y)+O((1-y)^2),&&y\to1,\label{horizon-y}
\end{align}
corresponding to the regular center and the similarity
horizon, respectively. The fundamental profile, found explicitly in \cite{B}, is
\begin{equation}\label{u0}
u_0(y)=\frac{1-y^2}{1+\frac{3}{5}y^2}.
\end{equation}
Numerical computations in \cite{BT} indicated that the
fundamental profile $u_0$ acts as a universal attractor
for generic blowup, while the first excited profile $u_1$
governs the threshold between dispersion and blowup.
The nonlinear asymptotic stability of $u_0$ under small
equivariant perturbations was subsequently established
in \cite{D,CDGH}.

Further numerical computations revealed a remarkable feature:
the least unstable mode has the same eigenvalue
$\lambda=5$ for every excited profile~\cite{B}, raising the question
of its structural origin.
 In this note we resolve this little spectral puzzle by constructing
the corresponding eigenfunction explicitly  in terms
of the profile.

\subsection*{Linearized eigenvalue problem}

Substituting
$
w(\tau,y)=u_n(y)+\varepsilon e^{\lambda\tau}v(y)
$
into \eqref{similarity} and retaining terms linear in
$\varepsilon$ yields the eigenvalue equation
\begin{equation}\label{mode-y}
(1-y^2)v''
+\left(\frac2y-2(\lambda+1)y\right)v'
+\left[
\frac{3(1-3u_n^2)}{y^2}-\lambda(\lambda+1)
\right]v=0.
\end{equation}
We seek eigenfunctions $v\in C^\infty([0,1])$.

One such eigenfunction arises from differentiating the
family of self-similar solutions with respect to $T$:
\[
\partial_T u_n\!\left(\frac{r}{T-t}\right)
=
-e^\tau y u_n'(y),
\]
which gives the gauge mode
\begin{equation}\label{gauge-y}
 v^{(n)}_g(y)=-yu_n'(y)\qquad\mbox{with}\qquad \lambda=1.
\end{equation}
It changes only the blowup time and is therefore not
counted as a physical instability.

\subsection*{Self-adjoint formulation}

For the spectral analysis and the exact-mode construction,
it is convenient to introduce the reciprocal coordinate
\[
x=\frac1y,\qquad  U(x)=u(y),\qquad V(x)=v(y),\qquad 1\le x<\infty.
\]
In these variables, equation \eqref{profile-y}
becomes
\begin{equation}\label{profile}
(x^2-1)U''+3U(1-U^2)=0.
\end{equation}
At the similarity horizon $x=1$ and the center $x=\infty$,
the profiles have the endpoint expansions
\begin{align}
U_n(x)&=b_n(x-1)-\frac{3b_n}{4}(x-1)^2
+\frac{5b_n}{16}(x-1)^3+O((x-1)^4),
&&x\to1,\label{left-expansion}\\
U_n(x)&=1-\frac{a_n}{x^2}+O(x^{-4}),
&&x\to\infty.\label{right-expansion}
\end{align}
The fundamental profile
\eqref{u0} becomes
\begin{equation}\label{U0}
U_0(x)=\frac{x^2-1}{x^2+\frac{3}{5}}.
\end{equation}
Equation \eqref{mode-y} is transformed into
\begin{equation}\label{mode-x}
pV''+\frac{2\lambda}{x}V'
+\left[3(1-3U_n^2)-\frac{\lambda(\lambda+1)}{x^2}\right]V=0,
\qquad p=x^2-1.
\end{equation}
Substituting
\begin{equation}\label{transformation}
V=x^{\lambda}p^{\frac{1-\lambda}{2}} \phi
\end{equation}
into~\eqref{mode-x} we obtain the self-adjoint Sturm--Liouville equation for  $\phi\in L^2((1,\infty),p^{-1}\,dx)$:
\begin{equation}\label{SL}
-(p\phi')'-3(1-3U_n^2)\phi=E p^{-1} \phi, \qquad E=-(\lambda-1)^2.
\end{equation}
Hereafter we  refer to $E$ as the energy and to $\lambda$ as the
similarity eigenvalue.
Although the transformation \eqref{transformation} relates
the two differential equations for every $\lambda$, it
identifies smooth similarity modes with square-integrable
Sturm--Liouville eigenfunctions only when
$\operatorname{Re}(\lambda)>1$. To see this, consider a smooth
mode with $V(1)\ne0$. Near $x=1$, its transform satisfies
\[
\phi\sim C(x-1)^{(\lambda-1)/2},
\]
which is square integrable with weight $p^{-1}$ precisely
for $\operatorname{Re}(\lambda)>1$. In that half-plane the
second branch is not square integrable. At infinity, the regular branch $V=O(x^{-2})$
corresponds to $\phi=O(x^{-3})$, whereas the singular
branch $V\sim Cx^3$ corresponds to $\phi\sim Cx^2$.
Only the regular branch is square integrable with
weight $p^{-1}$ at infinity.

Thus, for $\operatorname{Re}(\lambda)>1$, the similarity
eigenmodes of~\eqref{mode-x} are in one-to-one correspondence
with the eigenfunctions of the 
problem~\eqref{SL}, allowing us to apply  Sturm--Liouville theory.
 In particular, the self-adjointness of the Sturm--Liouville operator
implies that $E$ is real and hence that $\lambda$ is real.
The analogous correspondence for wave maps is discussed
in~\cite{BBM}. This equivalence does not apply to similarity modes with
$\operatorname{Re}(\lambda)<1$.

The gauge mode \eqref{gauge-y} becomes
$
V^{(n)}_g(x)=xU_n'(x),
$
which under~\eqref{transformation} gives
the zero-energy resonance (a bounded but
not square-integrable solution)
\[
\phi^{(n)}_g=U_n'.
\]
From the proof of existence  in~\cite{B}, it follows that the profile $U_n$ and its derivative each have exactly~$n$ interior zeros, which interlace.
Therefore, the Sturm oscillation theorem implies that the spectral
problem \eqref{SL} has exactly $n$ negative energy levels. We
order them as
\begin{equation}\label{energies}
0>E_1^{(n)}>\cdots>E_n^{(n)},
\end{equation}
where $E_k^{(n)}$, $k=1,\ldots,n$, denote the negative energy levels
for the profile $U_n$. 
The corresponding
eigenfunctions are denoted by $\phi_k^{(n)}(x)$.
The associated unstable similarity eigenvalues are ordered as
\begin{equation}\label{lambdas}
1<\lambda_1^{(n)}<\cdots<\lambda_n^{(n)},
\end{equation}
with the corresponding  eigenfunctions denoted by $V_k^{(n)}(x)=v_k^{(n)}(y)$. 
\vskip 0.1cm

\begin{table}[ht]
\centering
\begin{tabular}{c|ccccc}
$n$ & $\lambda_1^{(n)}$ & $\lambda_2^{(n)}$
    & $\lambda_3^{(n)}$ & $\lambda_4^{(n)}$
    & $\lambda_5^{(n)}$ \\
\hline
1 & 5 & --- & --- & --- & --- \\
2 & 5 & 28.407078 & --- & --- & --- \\
3 & 5 & 28.378841 & 183.492964 & --- & --- \\
4 & 5 & 28.374054 & 183.179999 & 1214.561254 & --- \\
5 & 5 & 28.373323 & 183.129311 & 1212.339587 & 8070.020343 \\
\hline
$\infty$ & 5 & 28.373193 & 183.120247 & 1211.917424 & 8052.273337
\end{tabular}
\caption{Unstable similarity eigenvalues $\lambda_k^{(n)}$ for $n=1,\ldots,5$
and their fixed-$k$ limits as $n\to\infty$.
The limiting row is obtained from
\eqref{limiting-quantization}.}
\label{tab:spectra}
\end{table}

Table~\ref{tab:spectra} presents the numerically computed
eigenvalues for the first five excited profiles, together with
their limiting values as $n\to\infty$. The striking feature of these numerical results, first observed
in~\cite{B}, is that the least unstable eigenvalue
$\lambda_1^{(n)}=5$ is common to all excited profiles,
whereas the remaining eigenvalues depend on $n$.
Equivalently, every excited profile shares the same highest
negative energy level $E_1^{(n)}=-16$.

The limiting values
$\lambda_k^{(\infty)}=\lim_{n\to\infty}\lambda_k^{(n)}$
were determined in~\cite{B} by selecting an appropriate
self-adjoint extension of the limiting Schr\"odinger operator. Writing $\lambda_k^{(\infty)}=1+\alpha_k$, they satisfy
\begin{equation}\label{limiting-quantization}
\operatorname{Im}\left[
\log\Gamma\left(\frac12+\alpha_k+\frac{i\sqrt{11}}2\right)
-\log\Gamma\left(\frac12+\frac{i\sqrt{11}}2\right)
\right]=k\pi,
\qquad k=1,2,\ldots,
\end{equation}
where $\log\Gamma$ denotes the analytic branch that is real
on the positive real axis.
In particular, $\alpha_1=4$ is an exact root, giving
$\lambda_1^{(\infty)}=5$. However, this limiting result leaves open why
$\lambda_1^{(n)}=5$ holds exactly for every finite $n\ge1$.
We now resolve this question by constructing the associated
eigenfunction explicitly and showing that it corresponds to
the highest negative energy level.

\subsection*{Construction of the exact mode}

The same energy level for every excited
profile suggests seeking a corresponding eigenfunction
given
by a universal first-order expression in the profile (here we suppress the index $n$ on $U_n$):
\begin{equation}\label{first-order-ansatz}
\phi=A(x)U'+B(x)U,
\end{equation}
where $A$ and $B$ are independent of the particular
profile $U$. The gauge mode $\phi_g=U'$ is the
simplest example of such an expression. We now seek
a nonzero energy for which an ansatz of this
form solves the linearized equation for every profile.

Substituting the ansatz~\eqref{first-order-ansatz} into equation~\eqref{SL} and using the profile equation~\eqref{profile} and its derivative to eliminate
$U''$ and $U'''$, we obtain
\begin{equation}\label{coefficient-identity}
\begin{aligned}
&\left(
pA''+p'A'+2pB'+p'B+\frac{E}{p}A
\right)U'
+
\left(
pB''+p'B'+\frac{E}{p}B-6A'
\right)U \\
&\qquad {}+6(A'-B)U^3=0.
\end{aligned}
\end{equation}
A sufficient condition for this expression to vanish
for every profile is that the three coefficients
vanish separately. Setting the coefficient of the cubic
term to zero  gives $B=A'$,
so the ansatz~\eqref{first-order-ansatz} takes the total-derivative form
$
\phi=(AU)',
$
while the remaining coefficient equations reduce to 
\begin{equation}\label{A-equations}
3pA''+2p'A'+\frac{E}{p}A=0,\qquad
pA'''+p'A''
+\left(\frac{E}{p}-6\right)A'=0.
\end{equation}
Taking $A$ to be constant recovers the gauge mode with $E=0$. To find a nonconstant solution, we seek $A$ bounded at
infinity and allow a simple pole at $x=1$, where every
profile vanishes. The simplest rational ansatz in
$p=x^2-1$ with these properties is
\[
A=a+\frac{b}{p},\qquad b\ne0.
\]
Substituting this into \eqref{A-equations} and clearing
denominators gives
\[
(Ea+10b)p+(E+16)b=0,
\qquad
2bx(E+16)=0.
\]
Since $b\ne0$, the second equation yields $E=-16$,
and then the first one gives $a=5b/8$.
Normalizing 
$a=1$, we obtain
\[
A=1+\frac{8}{5p}
=\frac{5x^2+3}{5(x^2-1)}
=\frac1{U_0}.
\]
Consequently, for every excited profile $U_n$, $n\ge1$,
the first-order ansatz yields the mode, provisionally denoted by $\phi_n$,
\begin{equation}\label{phi5}
\boxed{\phi_n=\left(\frac{U_n}{U_0}\right)'},
\end{equation}
which satisfies the eigenvalue equation~\eqref{SL} with $E=-16$,
corresponding to $\lambda=5$.

It remains
to verify the behavior of $\phi_n$ at the endpoints.
The endpoint expansions \eqref{left-expansion}
and \eqref{right-expansion} give
\[
\phi_n=O\bigl((x-1)^2\bigr)
\quad\text{as }x\to1,
\qquad
\phi_n=O(x^{-3})
\quad\text{as }x\to\infty.
\]
Hence $\phi_n\in L^2\bigl((1,\infty),p^{-1}\,dx\bigr)$,
completing the proof that $\phi_n$ is an eigenfunction
of the spectral problem \eqref{SL} with energy $E=-16$ for every $n\geq 1$.

To locate this energy level in the spectrum, let us count
the nodes of $\phi_n$. Since $U_0>0$ on $(1,\infty)$,
the quotient $U_n/U_0$ has exactly $n$ zeros. Therefore,
Rolle's theorem  gives at least $n-1$ distinct
interior zeros of $\phi_n$. On the other hand, the spectral
problem \eqref{SL} has exactly $n$ negative energy levels,
so the Sturm oscillation theorem allows at most $n-1$
interior zeros for an eigenfunction of negative energy.
Thus $\phi_n$ has exactly $n-1$ nodes, and $E=-16$ is
the highest negative energy level. In the notation~\eqref{energies},  this reads
\begin{equation}\label{phi1}
E_1^{(n)}=-16,\qquad
\phi_1^{(n)}(x)
=\left(\frac{U_n(x)}{U_0(x)}\right)'.
\end{equation}
The corresponding similarity eigenfunction with $\lambda_1^{(n)}=5$, expressed in the original
coordinate $y$, is
\begin{equation}\label{v1}
v_1^{(n)}(y)
=-\frac{y}{5(1-y^2)^4}
\left[(5+3y^2)(1-y^2)u_n'(y)+16y\,u_n(y)\right].
\end{equation}

\begin{remark*}
We emphasize that, although the linearized operators
share the eigenvalue $E_1^{(n)}=-16$, they do not form an
isospectral hierarchy. 
As Table~\ref{tab:spectra} illustrates, the remaining discrete
eigenvalues depend on $n$: increasing $n$ introduces an
additional bound state below $E=-16$, but also shifts the
other levels. This differs from a Darboux transformation
that creates a bound state while preserving the existing
bound-state energies. Our construction provides an exact
eigenfunction at one distinguished energy level directly in terms
of each nonlinear profile, rather than an intertwining map
between the linearized operators associated with different
profiles.
\end{remark*}

\subsection*{An underlying differential identity}

It is instructive to recast the preceding calculation
without imposing the profile equation. Writing
\[
\mathcal E[U]=pU''+3U(1-U^2),
\qquad
\mathcal L_{U,E}
=p\partial_x^2+p'\partial_x+3-9U^2+\frac{E}{p},
\]
we obtain, for arbitrary sufficiently smooth functions
$A$ and $U$,
\[
\begin{aligned}
\mathcal L_{U,E}\bigl[(AU)'\bigr]
&=(A\partial_x+3A')\mathcal E[U] \\
&\quad {}+\left(3pA''+2p'A'+\frac{E}{p}A\right)U'
+\left(pA'''+p'A''
+\left(\frac{E}{p}-6\right)A'\right)U.
\end{aligned}
\]
With $A=U_0^{-1}$ and $E=-16$, the last two terms
vanish by \eqref{A-equations}, yielding the identity
\begin{equation}\label{residual-identity}
\boxed{
\mathcal L_{U,-16}\left[\left(\frac{U}{U_0}\right)'\right]
=
U_0^{-1}\left(\partial_x
-\frac{3U_0'}{U_0}\right)\mathcal E[U],
}
\end{equation}
which is valid for any sufficiently smooth function $U(x)$.
In particular, for every excited profile $U_n$, the identity
shows that $(U_n/U_0)'$ satisfies the eigenvalue equation
with $E=-16$, recovering the exact mode constructed above. Whether this identity reflects a deeper structure  is unclear to us.

\subsection*{Special modes in $d=7$ and $d=9$}
A natural question is whether  similar special
eigenvalues occurs in higher dimensions.
Below we repeat the argument given in~[5] for wave maps
in higher odd dimensions.
In dimension $d$, the change of variables
\begin{equation}\label{SLd}
v(y)=y^{-\frac{d-3}{2}}
(1-y^2)^{\frac{d-3-2\lambda}{4}}\phi(x),
\qquad x=\frac1y,
\end{equation}
transforms the similarity-mode equation to a Sturm--Liouville
equation analogous to~\eqref{SL} with
\[
E=-\left(\lambda-\frac{d-3}{2}\right)^2.
\]
The gauge eigenvalue $\lambda=1$ and its reflection
$\tilde\lambda=d-4$ give the same energy
$E=-(d-5)^2/4$.
Thus, the same solution $\phi$ that gives the gauge mode
$v_g=-yu'$ for $\lambda=1$ through~\eqref{SLd} 
gives the formal reflected mode
\[
\tilde v_g=-\frac{yu'}{(1-y^2)^{\frac{d-5}{2}}}\qquad\mbox{with}\qquad \tilde\lambda=d-4.
\]
For $d=5$, the reflection fixes $\tilde\lambda=\lambda=1$, so it yields no new mode.
However, for $d=7$ and $d=9$ there exist infinitely many  smooth profiles $u_n$
that vanish at $y=1$ and satisfy
\[
u_n(y)\sim b_n(1-y)^{\frac{d-3}{2}},
\]
so their reflected
modes $\tilde v_g^{(n)}$ are smooth at $y=1$, and hence admissible.
Therefore all such
profiles share the eigenvalue $\tilde\lambda=3$
in $d=7$ and $\tilde\lambda=5$ in $d=9$,
corresponding to the energies $E=-1$ and $E=-4$,
respectively.

The above reflected  modes in $d=7$ and $d=9$ can also
be obtained directly from the first-order
ansatz~\eqref{first-order-ansatz}, which in these cases
reduces to the derivative term alone ($B=0$).
In contrast, for the case $d=5$ considered here, the
ansatz produces a mode that reflection does not.

\enlargethispage{\baselineskip}

\end{document}